\documentstyle[twoside,fleqn,espcrc2,epsf]{article}

\newcommand{\AmS}{{\protect\the\textfont2
  A\kern-.1667em\lower.5ex\hbox{M}\kern-.125emS}}
\def\ltapprox{\mathrel{{\lower 3pt\hbox{$\mathchar"218$}}
 \!\!\!\!\!\raise 2.0pt\hbox{$\mathchar"13C$}}}

\title{
       {\vspace{-2cm} \normalsize
       \hfill \parbox{21mm}{BI-TP 99/33}}\\[18mm]
       Numerical evidence for Goldstone-mode effects in the 
       three-dimensional $O(4)$-model\thanks{Talk presented by Tereza Mendes}}

\author{J\"urgen Engels and Tereza Mendes\address{Fakult\"at f\"ur Physik,
        Universit\"at Bielefeld, D-33615 Bielefeld, Germany}}
\begin{document}

\begin{abstract}
We investigate the three-dimensional $O(4)$-model on $24^3$--$96^3$
lattices as a function of the magnetic field $H$. Below the critical 
point, in the broken phase, we confirm explicitly the $H^{1/2}$ 
dependence of the magnetization and the corresponding $H^{-1/2}$
divergence of the longitudinal susceptibility, which are predicted from
the existence of Goldstone modes. At the critical point the magnetization 
follows the expected finite-size-scaling behavior,
with critical exponent $\delta = 4.87(1)$.
\end{abstract}

\thispagestyle{empty}
\maketitle

The continuous symmetry present in the $O(N)$ spin models
$$
\beta\,{\cal H}\;=\;-J \,\sum_{<i,j>} {\bf S}_i\cdot {\bf S}_j 
         \;-\; {\bf H}\cdot\,\sum_{i} {\bf S}_i
$$
--- where ${\bf S}_i$ are unit vectors in an $N$-dimensional 
sphere --- gives rise to the so-called {\em spin waves}: slowly varying
(long-wavelength) spin configurations, whose energy may be arbitrarily
close to the ground-state energy. In two dimensions these modes are
responsible for the absence of spontaneous magnetization,
whereas in $d>2$ they are the Goldstone modes associated with the 
spontaneous breaking of the rotational symmetry for
temperatures below the critical value $T_c$. Due to the presence
of Goldstone modes, a diverging susceptibility is induced
in the limit of small external magnetic field $H$ for all $T<T_c$,
i.e.\ everywhere on the coexistence curve (see for example \cite{WZ}). 
More precisely: not only the {\bf transverse susceptibility}, which is 
directly related to the fluctuation of the transverse (Goldstone) modes, 
diverges when $H\to0$ 
$$\chi_T(T<T_c,H) \;=\; \frac{M(T,H)}{H} \;=\; {\cal O}(H^{-1})$$
but also the {\bf longitudinal susceptibility},
given in terms of the magnetization $M$ by $\chi_L \equiv \partial M/
\partial H\;,$ is predicted to diverge for $2<d\leq4$. In
$d=3$ the predicted divergence is $H^{-1/2}$, or
equivalently, the behavior for the magnetization will include a
``singular'' term of order $H^{1/2}$
$$M(T<T_c,H)\;=\;M(T,0)\,+\,c\,H^{1/2}\;.$$
Indication of this behavior was seen in early numerical simulations 
of the three-dimensional $O(3)$ model at low temperatures \cite{MK}.

Here we verify explicitly the predicted singularities --- or Goldstone-mode 
effects --- for the three-dimensional $O(4)$ case, by simulating the model
in the presence of an external magnetic field and close to the critical 
temperature $T_c$. The low-temperature singularities are still present at 
$T\ltapprox T_c$. We are able to compute magnetic critical exponents 
directly, by varying $H$ at $T_c$, and we also {\em use} the observed 
Goldstone-effect behavior to extrapolate our data to $H\to0$ and obtain 
the zero-field critical exponents in very good agreement with the existing 
values. Using these exponents, we verify finite-size scaling at $T_c$.
Our motivation for considering the three-dimensional $O(4)$ model 
is that it is expected to be in the same 
universality class as QCD at finite temperature for two
degenerate quark flavors \cite{RW}, the magnetic field
in the spin model corresponding to the quark mass in QCD. 

Our simulations are done using the cluster algorithm.
We compute the magnetization along the direction of
the magnetic field as
$${M}\;=\; %-\partial F/\partial H \;=\; 
     \frac{1}{V}<\!\sum_{i} {\bf S}_i\!\cdot\!{\bf \hat H}\!>\;. $$
We note that, due to the presence of a nonzero field, the magnetization
defined above is nonzero on finite lattices, contrary to what happens
for simulations at $H=0$, where one is led to consider 
$1/V <\!|\sum {\bf S}_i|\!>$, which overestimates the true magnetization.

The susceptibilities are obtained from the spin correlation
functions
\begin{eqnarray}
G_L(x)&\equiv&<\!S_0^{\parallel}S_x^{\parallel}\!> - 
                       <\!S_0^{\parallel}\!><\!S_x^{\parallel}\!> 
\nonumber \\
G_T(x)&\equiv&\frac{1}{3}<\!{\bf S}_0^{\perp}\cdot{\bf S}_x^{\perp}\!>
\nonumber
\end{eqnarray}
by
$
{\chi_L}\,=\,\partial M/\partial H\,=\,{\widetilde G_L}(0)\,;\;
{\chi_T}\,=\,{\widetilde G_T}(0)\;.
$
We use for $T_c$ the value 
$$1/T_c \;\equiv\; J_c \;=\; 0.93590 \;, $$
obtained in simulations of the zero-field model
\cite{H0}. 

\begin{figure}
\vspace{-2.6cm}
%\leavevmode
\epsfxsize = 0.63\textwidth
\centerline{\hskip 1.02cm \epsffile{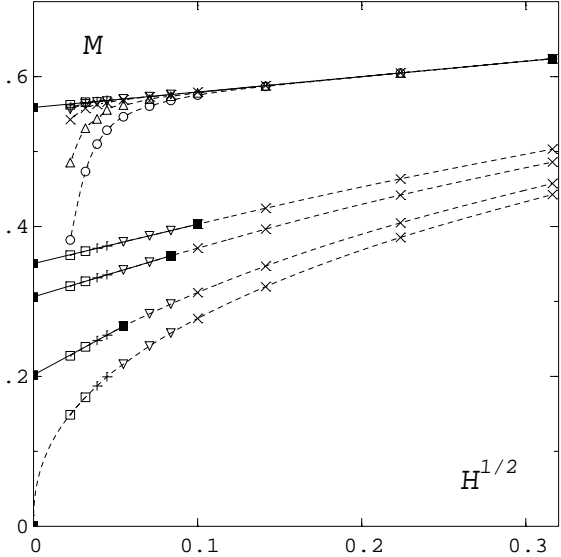}}
\vspace{-4.6cm}
\caption{Magnetization for $T\leq T_c$.}
\label{fig1}
\vspace{-0.5cm}
\end{figure}
%\begin{figure}
%\vspace{-3cm}
%\epsfxsize = 0.5\textwidth
%\leavevmode\epsffile{SEMINAR/mj.ps}
%\vspace{-3cm}
%\caption{Extrapolation $H \to 0$}
%\end{figure}

\begin{figure}
\vspace{-2.6cm}
\epsfxsize = 0.63\textwidth
\centerline{\hskip 10mm\leavevmode\epsffile{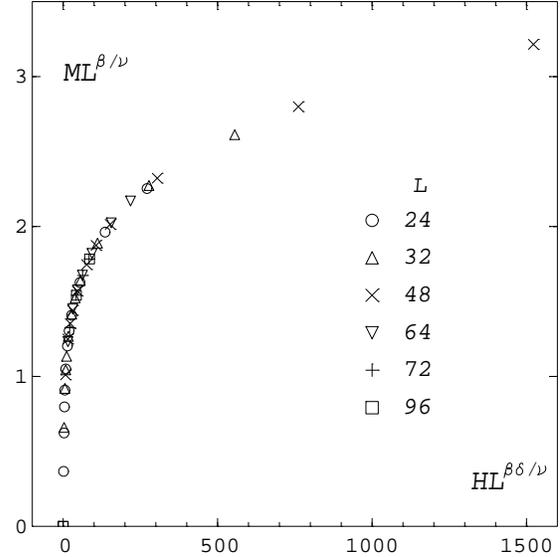}}
\vspace{-4.65cm}
\caption{Finite-size-scaling plot for the magnetization at $T_c$.}
\label{FSS}
\vspace{-0.45cm}
\end{figure}

In Fig.\ \ref{fig1} we show our data for the magnetization for
temperatures $T\leq T_c$ plotted versus $H^{1/2}$. Inverse temperatures
are given by $J=J_c$, 0.95, 0.98, 1.0, 1.2, starting from the lowest 
curve. We have simulated at increasingly larger values of $L$ at fixed
values of $J$ and $H$ in order to eliminate finite-size effects.
Except for the curve at $J=1.2$ only these largest-lattice values are
shown. Symbols are as explained in Fig.\ \ref{FSS}.

The finite-size effects for small $H$ do not disappear as one moves
away from $T_c$, but rather increase.
Solid lines connect the $y$-axis to the last point (a filled square)
included in our fits of the data for small $H$. It can clearly be 
seen that the predicted behavior (linear in $H^{1/2}$) holds close 
to $H=0$ for all temperatures $T<T_c$ considered. We thus see evidence 
that the Goldstone-mode effects are observable even very close to $T_c$. 
The straight-line segments become shorter as $T_c$ is approached from below, 
and at $T_c$ the magnetization vanishes as a power of $H$, as expected.

\begin{figure}
\vspace{-2.6cm}
\epsfxsize = 0.6\textwidth
\centerline{\hskip 12mm \leavevmode\epsffile{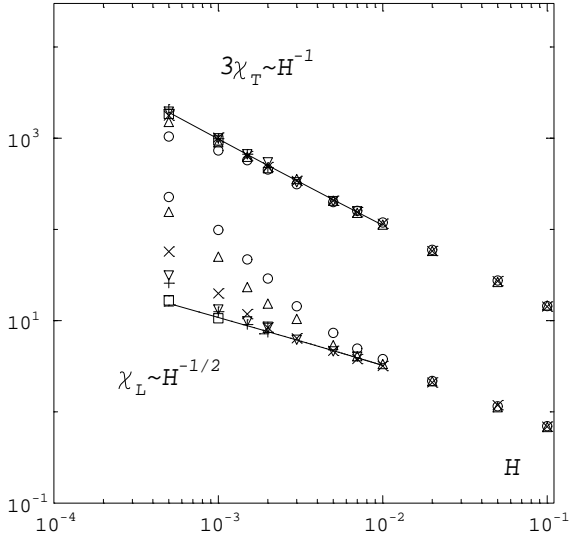}}
\vspace{-4.5cm}
\caption{Susceptibilities $\chi_T$ and $\chi_L$ at low $T$.}
\label{chi_low}
\vspace{-0.5cm}
\end{figure}

We have fitted the data from the largest
lattice sizes at $T_c$ to the scaling behavior
$$M(T_c,H) \;\sim\; H^{1/\delta},$$
obtaining the exponent $\delta=4.87(1)$,
in agreement with \cite{H0}. The straight-line fits shown in 
Fig.\ \ref{fig1} are used to extrapolate our data to the zero-field 
limit, and the corresponding zero-field scaling law
$$M(T\ltapprox T_c,H=0) \;\sim\; (T_c - T)^{\beta} $$
yields $\beta=0.38(1)$, also in agreement with \cite{H0}.
The finite-size effects at $T_c$ are very well described by
the finite-size-scaling ansatz
$$ M(T_c,H;L) \,=\, L^{-\beta/\nu}\, Q_M(H\,L^{\beta\delta/\nu})\;,$$
as shown in Fig.\ \ref{FSS}. We see no corrections to finite-size
scaling.
In Fig.\ \ref{chi_low} we show a typical behavior for $\chi_T$ and
$\chi_L$ at low temperature, for $J\;=\;0.98$.
Symbols are as in Fig.\ \ref{FSS}.
The lines $M/H$ (a factor 3 is included for clarity) and 
$\partial M/\partial H$ (from the fit in Fig.\ \ref{fig1}) are shown, 
for comparison respectively with $\chi_T$ and $\chi_L$. 
Both lines are obtained from our ``infinite-volume'' data.
Notice the large finite-size effects in $\chi_L$.

%\begin{figure}
%\vspace{-3cm}
%\epsfxsize = 0.5\textwidth
%\leavevmode\epsffile{SEMINAR/csi_lowT.ps}
%\vspace{-3cm}
%\end{figure}

\begin{figure}
\vspace*{-2.3cm}
\epsfxsize = 0.55\textwidth
\centerline{\hskip 7mm\leavevmode\epsffile{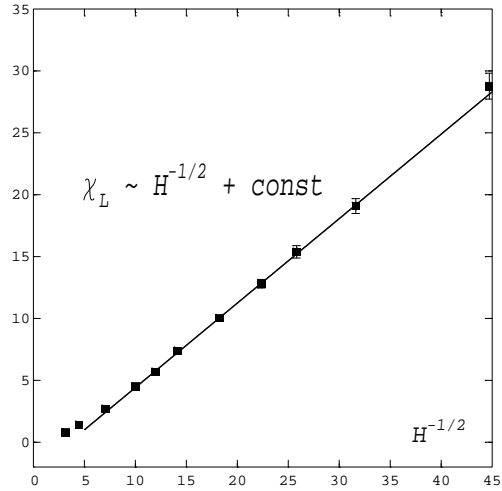}}
\vspace*{-4.3cm}
\caption{The susceptibility $\chi_L$ at $J=0.95$.}
\label{gold}
\vspace*{-0.7cm}
\end{figure}

\begin{figure}
\vspace*{-2.3cm}
\epsfxsize = 0.55\textwidth
\centerline{\hskip 7mm\leavevmode\epsffile{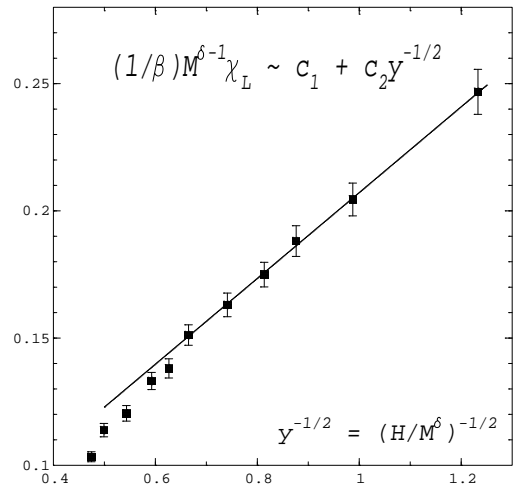}}
\vspace*{-4.3cm}
\caption{$(1/\beta) M^{\delta - 1} {\chi_L}$ at $J=0.95$.}
\label{zia}
\vspace*{-0.5cm}
\end{figure}

Finally, we verify numerically the claim in Ref.\ \cite{WZ},
that the Goldstone-mode singularity at low temperatures is consistent
with the magnetic equation of state
$$ (1/\beta) M^{\delta - 1} {\chi_L} \approx
c_1 \,+\, c_2 \, {y}^{-1/2}\;, $$
where ${y}\equiv {H}/M^{\delta}$. In Figs.\ \ref{gold} and \ref{zia}
we show respectively the behavior of $\chi_L$ and 
the combination above for $J=0.95$.
We see that the two perturbative predictions are well satisfied at 
$T\ltapprox T_c$. %$J=0.95$. 
Note that the line in Fig.\ \ref{zia}
is not obtained using the perturbative coefficients given in \cite{WZ},
but from a fit of our data. This idea of finding nonperturbative 
coefficients from fits to perturbative forms is very useful in
obtaining an expression for the scaling function for this model
\cite{scafun}.

\vskip 2mm
This work was supported by the DFG under Grant No.\ Ka 1198/4-1 and
by the TMR-Network ERBFMRX-CT-970122.

\protect\vspace{-6.5cm}

\end{document}